\documentstyle[prl,twocolumn,aps,epsf]{revtex}

\newcommand{\extraspace}{\addtolength{\abovedisplayskip}{1mm} 
                        \addtolength{\belowdisplayskip}{1mm} 
                        \addtolength{\abovedisplayshortskip}{1mm} 
                        \addtolength{\belowdisplayshortskip}{1mm}} 
\newcommand{\be}{\begin{equation}\extraspace} 
\newcommand{\ee}{\end{equation}} 
\newcommand{\bea}{\begin{eqnarray}\extraspace} 
\newcommand{\eea}{\end{eqnarray}} 
 
\newcommand{\nonu}{\nonumber \\[2mm]}

\newcommand{\half}{{\textstyle \frac{1}{2}}}

\newcommand{\eps}{\epsilon}

\newcommand{\up}{\uparrow} 
\newcommand{\down}{\downarrow}

\newcommand{\del}{\partial} 
 
\newcommand{\ie}{{\it i.e.}}

\begin{document} 
 
\title{Non-abelian Electrons}
\author{Peter Bouwknegt}
\address{ 
   Department of Physics and Mathematical Physics, 
   University of Adelaide, Adelaide, SA 5005, AUSTRALIA}
\author{Kareljan Schoutens} 
\address{ 
     Institute for Theoretical Physics,
     Valckenierstraat 65, 1018 XE Amsterdam, 
     THE NETHERLANDS} 
\date{May 18, 1998}
\maketitle 
\begin{abstract} 
We analyze critical and massive $SO(5)$ superspin 
regimes for correlated electrons on a two-chain ladder.
We identify fundamental low energy excitations,
which carry the quantum numbers of a free electron,
and can be probed in (inverse) photo-emission experiments.
These excitations do not obey the usual Pauli Principle, but 
are governed by specific forms of so-called non-abelian 
exclusion statistics. 

\end{abstract}

\vskip 1mm
\noindent{\small Report number: ADL-98-12/M64, ITFA-98-10}

\vskip 2mm

\narrowtext

It is well-known that the quantum statistics of particles in
one or two spatial dimensions are not necessarily bosonic or
fermionic. In two spatial dimensions all that is needed is
a representation of the braid group, and this allows for
various alternative possibilities. A particularly intriguing
possibility is that of so-called non-abelian braiding statistics,
in which case the braiding of particles is represented by 
non-trivial matrices acting on multi-component wave functions.
A prototype example for this are the so-called quasi-hole
excitations over the pfaffian quantum Hall state \cite{pfaf}.

In a recent paper \cite{Sc2}, one of us investigated the exclusion 
statistics properties of the edge quasi-holes over the pfaffian 
quantum Hall state. A key quantity is the thermodynamic distribution 
function $n_{\rm qh}(\eps)$, which describes the expected 
occupation of specific one-quasi-hole states and which generalizes
the Fermi-Dirac distribution for fermions. The edge quasi-hole
distribution, which was computed in \cite{Sc2}, has the 
following characteristic features
\bea
  n_{\rm qh}(\eps) &\sim&  8 \qquad \qquad {\rm for}\ \eps \ll \mu
\nonu
              &\sim&  \sqrt{2} \, e^{-\beta \eps} \quad 
                      {\rm for}\ \eps \gg \mu
\eea
with $\beta=(k_B T)^{-1}$ and $\mu$ the chemical potential.
The nontrivial prefactor $\sqrt{2}$ in the high-energy Boltzmann 
tail of the distribution $n_{\rm qh}(\eps)$ is a direct manifestation 
of the non-abelian nature of the braiding statistics of the 
quasi-holes over the pfaffian quantum Hall state \cite{Sc2}. 

One may demonstrate that, in general, non-abelian braiding
statistics in a Conformal Field Theory (CFT) lead to non-trivial 
prefactors in the Boltzmann tails of the corresponding thermodynamic 
distribution functions \cite{Sc2,BS}. We therefore propose the 
term `non-abelian exclusion statistics' for all forms of exclusion 
statistics that exhibit such non-trivial prefactors. 
In the theoretical literature, a number of examples, other than the 
pfaffian, have been discussed. These include spinons 
in $S\geq 1$ critical spin chains \cite{FS}, and generalized 
fermions in minimal models of CFT \cite{BS}. 

In this Letter we focus on situations where excitations
with the quantum numbers of free electrons obey specific
forms of non-abelian exclusion statistics. The concrete 
setting for this is the so-called $SO(5)$ superspin regime for 
correlated electrons on a two-chain ladder. The excitations,
which we call `non-abelian electrons', can be probed via (inverse) 
photo-emission and, in principle, the non-abelian nature of their 
exclusion statistics can be tested in experiments.

Before discussing the $SO(5)$ ladder models and the 
associated non-abelian electrons, we shall briefly discuss 
the exclusion statistics properties of the spinons and holons
of free or weakly interacting electrons in one dimension.
In a continuum description, free electrons in one dimension are 
described by a Conformal Field Theory (CFT), with the electron 
degrees of freedom represented by four real (Majorana) fermion 
fields
\be
 \left( \psi_{(\up,e)}, \psi_{(\down,e)}, 
        \psi_{(\down,-e)}, \psi_{(\up,-e)} \right) \ .
\label{psis}
\ee
These fields transform as a vector under the $SO(4)$ symmetry.
Through non-abelian bosonization, the free fermion CFT is equivalent 
to an $SO(4)$ Wess-Zumino-Witten theory with level $k=1$ 
($SO(4)_1$ WZW).
 
As soon as (weak, repulsive) interactions are introduced in a 
one-dimensional electron theory,  a separation of spin and 
charge sets in and the fundamental excitations are found to be 
spinons (carrying spin-$\half$ and no charge)
and holons (carrying charge $\pm e$ and no spin).
An explicit example is the $U>0$ Hubbard model, where one 
may use the exact Bethe Ansatz solution to demonstrate the 
fundamental role of the spinon and holon excitations \cite{EK}. 
The free electron CFT may equally well be described in terms 
of spinons and holons, which we write as
\be
(\phi_\up,\phi_\down) \  , \qquad 
 (\phi_e,\phi_{-e}) \ .
\ee 
Together, the spinon and holon fields form a spinor representation
of the $SO(4)$ symmetry. In figure 1 we show in an $SO(4)$ weight 
diagram the quantum numbers of the electron, spinon and holon 
fields.

This brings us to the statistics of the spinons and holons in a 
free electron theory. That these statistics are unusual 
can be demonstrated by the following argument. We consider the 
capacitance of the free electron gas, \ie, the amount of charge
$\Delta Q$  
pulled from the Fermi sea by an applied voltage $V$. In the
electron-operator picture, this response is generated by the 4 
fermionic fields, eq.~(\ref{psis}), of charge $\pm e$. In the 
spinon/holon picture, the response is generated by the holons 
alone, \ie, by 2 fields of charge $\pm e$. Assuming fermionic 
statistics for the holons thus leads to a mismatch by a factor of 
2 between the two approaches. As it turns out, the actual,
fractional, statistics of the holons are such that the response
to a voltage indeed exhibits an enhancement by a factor of 2 
w.r.t.\ free fermions (see below).

The appropriate framework for discusing the statistics of spinons 
and holons is the notion of `fractional exclusion statistics',
which was introduced by Haldane in \cite{Ha} and elaborated on 
in \cite{gstats}. The early applications of this new paradigm in 
many-body theory have hinged on the specifics of exactly solvable 
models of quantum mechanics with inverse square exchange. In an 
important step forward \cite{Sc1}, it was 
demonstrated that a fractional statistics assignment can be 
done as soon as an effective CFT for low-energy behavior has been 
identified. The method introduced in \cite{Sc1} is based on
recursion relations for truncated conformal spectra. In the mean 
time, a large number of examples of this 
`statistics from field theory' approach have been worked out
\cite{Sc1,ES,Sc2,FS,BS}.

To obtain the exclusion statistics of spinons and holons, we rely on 
our recent results, presented in \cite{BS}, for quasi-particle 
formulations of level-1 WZW models. For the case $SO(4)_1$, 
we obtained explicit results for the exclusion statistics of 
quasi-particles transforming in the spinor of $SO(4)$,  which  
consists of the spinon and holon fields combined. The result may 
be phrased in terms of a quantity $\lambda(x_s^\pm,x_c^\pm)$, 
which is the approximate partition sum for a single one-particle 
level in the spectrum. This partition sum depends on 
the one-particle energy $\eps$ and on chemical potentials 
$\mu_s^\pm$, $\mu_c^\pm$ for the various spinor components 
through the quantities $x_{s,c}^\pm = e^{\beta(\mu_{s,c}^\pm-\eps)}$.
In \cite{BS} we showed that
\be
\lambda(x_s^\pm,x_c^\pm) = \tilde{\lambda}(x_s^+,x_s^-) 
  \, \tilde{\lambda}(x_c^+,x_c^-) \ ,
\label{fac}
\ee
with the $\tilde{\lambda}(x^+,x^-)$ given by the single-level partition 
sum $\mu(x,z)$ for spinons in the $SU(2)_1$ WZW model \cite{Sc1},
\bea
\lefteqn{
\tilde{\lambda}(xz,xz^{-1}) = \mu(x,z) = }
\nonu
&& 1 + (z^2 + z^{-2}){x^2 \over 2}
   + (z+z^{-1}){x \over 2}\sqrt{(z-z^{-1})^2 x^2 + 4} \ .
\label{mu}   
\eea
The parameter $x$ is related to a chemical potential for spinons 
and $z$ keeps track of their $SU(2)$ quantum number $S^z$. The 
factorization (\ref{fac}) is a direct consequence of the identity 
$SO(4)_1 = SU(2)_1 \times SU(2)_1$  at the CFT level \cite{BPS}.
The exclusion statistics encoded in $\mu(x,z)$ agree with Haldane's 
fractional exclusion statistics,  with the statistics matrix given
by $G = \left( \begin{array}{cc} 1/2 & 1/2 
\\ 1/2 & 1/2 \end{array} \right)$. 

{}From eq. (\ref{fac}), (\ref{mu}), the spinon and holon thermodynamics 
are easily derived. The thermal response follows from 
$\mu(x,z=1) = (1+x)^2$,  leading to a CFT central charge $c_{CFT}=1$ 
for both spinons and holons. The response to a field follows from
\be
z\del_z \log[\mu(x,z)]
=
{2x(z-z^{-1}) \over
\sqrt{ (z-z^{-1})^2 x^2 + 4} } \ .
\ee
The above-mentioned factor of 2 in the holon response to a 
voltage is established by comparing the integral expressions
(with $\mu_c^\pm = \pm eV$ and $\rho$ the density of states)
\bea
&& \Delta Q_{\rm fermion}(\beta,V) =
\nonu
&& 
 \rho e \int_0^\infty d\eps \, 
{ e^{-\beta(\eps-eV)} - e^{-\beta(\eps+eV)}  \over
(1+e^{-\beta(\eps-eV)}) (1+e^{-\beta(\eps+eV)} )} = \rho e^2 V
\nonu
&& \Delta Q_{\rm holon}(\beta,V) =
\nonu
&&
 \rho e \int_0^\infty d\eps \, 
{2( e^{-\beta(\eps-eV)} - e^{-\beta(\eps+eV)} ) \over
\sqrt{ (e^{\beta e V} - e^{-\beta e V})^2 e^{-2\beta\eps} +4}}
= 2 \rho e^2 V 
\eea
for a single charged fermion and a holon, respectively. 

The low-temperature behavior of a one-dimensional electron gas 
(away from half-filling) in the presence of weak repulsive 
interactions is described by the Luttinger Liquid, which is 
parametrized by a single interaction parameter $K_c$ and two 
velocities $u_c$ and $u_s$ for gap-less excitations. 
The above discussion of the spinon and holon  statistics 
is easily generalized to the general Luttinger Liquid.
The absence of a non-trivial parameter $K_s$ makes clear that 
the spinon statistics in a Luttinger Liquid are identical to those 
in the free electron theory, while a non-trivial value $K_c\neq 1$
will lead to a modification of the holon statistics. 

We now turn to a specific systems of strongly correlated 
electrons, in which the fundamental degrees of freedom belong to 
the spinor representation of $SO(5)$, and are radically different 
from the spinons and holons in the weakly interacting case.
$SO(5)$ symmetry has recently been proposed as a structure that
unifies anti-ferromagnetic (AF) ordering and $d$-wave 
superconductivity (dSC). As such, it may be invoked in attempts to 
sort out the competition between AF and dSC ordering in the cuprate 
superconductors \cite{Zh}. As a laboratory for studying the
implications of $SO(5)$ symmetry, Scalapino, Zhang and Hanke (SZH)
\cite{SZH} have introduced a class of two-chain ladder models with 
exact $SO(5)$ symmetry \cite{SS}. In their paper, SZH give a 
strong-coupling phase diagram for these models, as a function of the 
couplings $U$, $V$ within a single rung. Among others, they identify 
a so-called $SO(5)$ superspin phase (called $E_1$), where the 
electron degrees of freedom on a single rung reduce to an $SO(5)$ 
vector $n_a(x)$, $a=1,\ldots,5$, composed of electron-bilinears that 
represent the AF and dSC order parameters,
\be
  n_a = \left( 
           {\Delta^\dagger+\Delta \over 2} , 
           N_x, N_y, N_z, 
           {\Delta^\dagger - \Delta \over 2i} 
        \right) \ .
\label{so5vec}
\ee

In the superspin phase, inter-rung interactions lead to an effective 
$SO(5)$ spin-chain, with the fundamental `spins' in the
vector-representation of $SO(5)$. This may be compared 
to an $SO(3)$ spin-chain with spins in the vector
representation, which is an $SU(2)$, $S=1$ spin-chain.
On the basis of this analogy, one expects that generic 
inter-rung couplings in an $SO(5)$ ladder lead to a `Haldane' 
gap. A particular example of a gapped phase (the so-called
AKLT point) was studied in \cite{SZH}. One also expects 
a single critical point, which avoids the Haldane gap, and
which is analogous to the integrable $S=1$ chain. It will be
represented by inter-rung interactions
\be
  \sum_j \left[  \alpha_1 (L^{ab}_j \, L^{ab}_{j+1}) 
+ \alpha_2 (L^{ab}_j \, L^{ab}_{j+1})^2 \right]
 \ee
with $L^{ab}$ the 10 generators of $SO(5)$ and the ratio 
$\alpha_2/\alpha_1$ tuned to a critical value. Deviations from 
the critical $\alpha_2/\alpha_1$ will be relevant and will destroy 
the critical behavior.

These qualitative expectations are in agreement with the 
perturbative RG analysis of \cite{LBF}, which has identified  
two $SO(5)$-invariant fixed points: a massless point, described
by the  $SO(5)_1$ WZW theory  and a massive point corresponding
to the so-called $SO(5)$ Gross-Neveu (GN)  field theory.

In what follows, we shall first focus on the critical $SO(5)_1$ WZW 
theory , where we establish the existence of non-abelian electrons and 
give full quantitative results for their exclusion statistics. After that, 
we consider the massive $SO(5)$ GN theory, and argue that also in this 
massive phase non-abelian electrons are present. We shall also explain 
how these excitations can be probed in photo-emission type experiments.

Through non-abelian bosonization, the critical $SO(5)_1$ WZW theory 
is equivalent to a theory of 5 free fermions $\psi_a$. 
These fermions carry the quantum numbers of the $SO(5)$ vector 
$n_a$, eq. (\ref{so5vec}), which was formed as a bilinear in the 
original electron operators on the ladder.

In close analogy to the spinon/holon formulation of the free electron
CFT, we here propose a formulation of the $SO(5)_1$ WZW 
theory in terms of fundamental quasi-particles $\phi$ that transform 
in the (4-dimensional) spinor representation of $SO(5)$. 
Inspecting the quantum numbers for spin and charge, one
finds (see figure 2) that the 4 spinor components carry the 
quantum numbers $S^z=\pm\half$, $q=\pm e$ of a free electron:
\be
\left(
\phi_{(\up,e)}, \phi_{(\down,e)}, \phi_{(\down,-e)}, \phi_{(\up,-e)}
\right) \ .
\ee

The exclusion statistics of spinor quasi-particles in the
$SO(5)_1$ WZW theory have been obtained in our paper
\cite{BS}. An important aspect of this analysis has been the 
fact that, group-theoretically, a product of two $SO(5)$ spinors
contains both the vector representation and the singlet.
This means that, for example, a state constructed from
the vacuum by exciting two $\phi$-quanta can be either
an $SO(5)$ singlet (\ie, a neutral, spin-less excitation),
or an $SO(5)$ vector carrying the quantum numbers of the
$SO(5)$ order parameter $n_a$. If we consider a 4-$\phi$
state, and insist that it forms an $SO(5)$ singlet,
we have contributions from two distinct `fusion channels'
depending on the nature of the intermediate state with the first 
two $\phi$-quanta in place. When building the full spectrum in 
terms of $\phi$-quanta, one has to specify a `fusion channel' 
for each multi-$\phi$ state, and this gives a degeneracy factor 
of 2 for every two $\phi$-quanta that are added. The 
geometric average of these degeneracy factors equals
$\sqrt{2}$ per $\phi$-quantum and this factor ends up in the 
Boltzmann tail of the generalized distribution for these 
quasi-particles. 

Summarizing the above, we see that the spinor quasi-particles 
in the $SO(5)_1$ theory, which carry the quantum numbers of a 
free electron, obey a form of non-abelian exclusion statistics
and are properly called `non-abelian electrons'.

In our paper \cite{BS}, we obtained quantitative results for
the exclusion statistics of $SO(5)_1$ spinor quasi-particles.
These results can be phrased in terms of a 
single-level partition sum $\lambda(x)$, with $x$ keeping 
track of the various chemical potentials and energy.
For the case where one gives equal weight to all
four components, $x=e^{\beta(\mu-\eps)}$ with
$\mu$ an overall chemical potential, $\lambda(x)$ 
is a solution of
\bea
\lefteqn{
\lambda^{3 \over 2}
-(2+3x^2) \lambda} 
\nonu
&& \quad 
+(3x^2-1)(x^2-1) \lambda^{1 \over 2} -x^2(x^2-1)^2  = 0 \ .
\label{poly}
\eea
The generalized Fermi-Dirac distribution is given by
\be
n_{\phi}(\eps) 
 = [x \del_x \log \lambda(x)](x=e^{\beta(\mu-\eps)}) \ . 
\ee
{}From (\ref{poly}) one immediately derives the properties
\be
\lambda(x) \sim x^4  \qquad {\rm for}\ x \gg 1 \ ,
\ee
which implies a maximum $n_{\phi}^{\rm max}=1$ for the 
occupation per level of each of the $\phi$-quanta, and
\be
\lambda(x) = 1 + 4\sqrt{2} \, x + {\cal O}(x^2) 
\qquad {\rm for}\ x \ll 1 \ ,
\ee 
which implies the prefactor $\sqrt{2}$ in the Boltzmann
tail of  $n_{\phi}(\eps)$.
 
Comparing the central charge $c^{CFT}$ and the
spin and charge susceptibilities $\chi^s$ and $\chi^c$
to those of standard free electrons (eq. (\ref{psis})), 
one finds
\be
c^{CFT}_\phi / c^{CFT}_e = 5/4 \ , \quad
\chi^s_\phi / \chi^s_e = \chi^c_\phi / \chi^c_e = 2 \ .
\ee 
These results are readily checked in a picture with 5 fundamental 
fermions. Note that the deviations from the free electron results
are not fermi-liquid effects, but manifestations of  the 
unusual statistics of the non-abelian electrons. 
 
In the massive $SO(5)$ GN model, the $SO(5)$ vector and spinor
excitations exist in the form of massive particles with mass ratio
\cite{GN}
\be
M_{\psi} = \sqrt{3} \, M_{\phi} \ .
\ee
The $\phi$-particles, which are the lightest particles in the theory,
may be viewed as `kinks' that carry a topological charge related to
the $SO(5)$ group theory \cite{GN}. In principle, one may
derive the thermodynamics of the $\phi$-particles by starting
from an exact scattering matrix and applying the Thermodynamic 
Bethe Ansatz. To our knowledge this procedure has not been
carried out, but it is clear that also in this massive theory 
a Boltzmann tail prefactor equal to $\sqrt{2}$ will arise. 
[The detailed form of the distribution $n_{\phi}$ will differ between 
the critical and the massive cases.] We conclude that also in the 
massive case, the terminology `non-abelian electrons' is justified.

The physical processes by which non-abelian electrons
can be excited inside the $SO(5)$ ladders, both in the
critical and in the massive phase, are easily described
and can possibly be realized in experiments.
The physical set-up is a photo-emission process, where a high 
energy photon ejects an electron from the ladder. After 
such a process, or its inverse, one of the rungs in the ladder
violates the $SO(5)$ superspin condition, and the ladder finds 
itself in an excited state with the quantum numbers of a free 
electron. As we discussed, such excited states correspond to the 
$\phi$-quanta, \ie, to the non-abelian electrons.

Before concluding, we would like to comment on the 
the massive $SO(8)$ phase that was found in the perturbative 
analysis of the $SO(5)$ ladder models \cite{LBF}. This
phase features an $SO(8)$ vector of order parameters, 
bilinear in the bare electron operators, and $SO(8)$ spinors
that carry the quantum numbers of free electrons.
It is important to realize that the group theory of
$SO(8)$ works out in such a way that there are no non-trivial 
fusion channels in multiple tensor products of the spinor 
particles, and that as a consequence the phenomenon of 
non-abelian statistics does not occur for $SO(8)$.

The authors thank A.~de Visser, S.~Guruswamy, B.~Nachtergaele,  
S.~Sachdev,  J.~Zaanen and F.~Wilczek for discussions. This 
research was supported in part by the Australian Research Council 
and by the foundation FOM of the Netherlands.

\vskip 6mm

\vbox{
\epsfxsize=5.9cm
\epsfysize=4.1cm
\epsffile{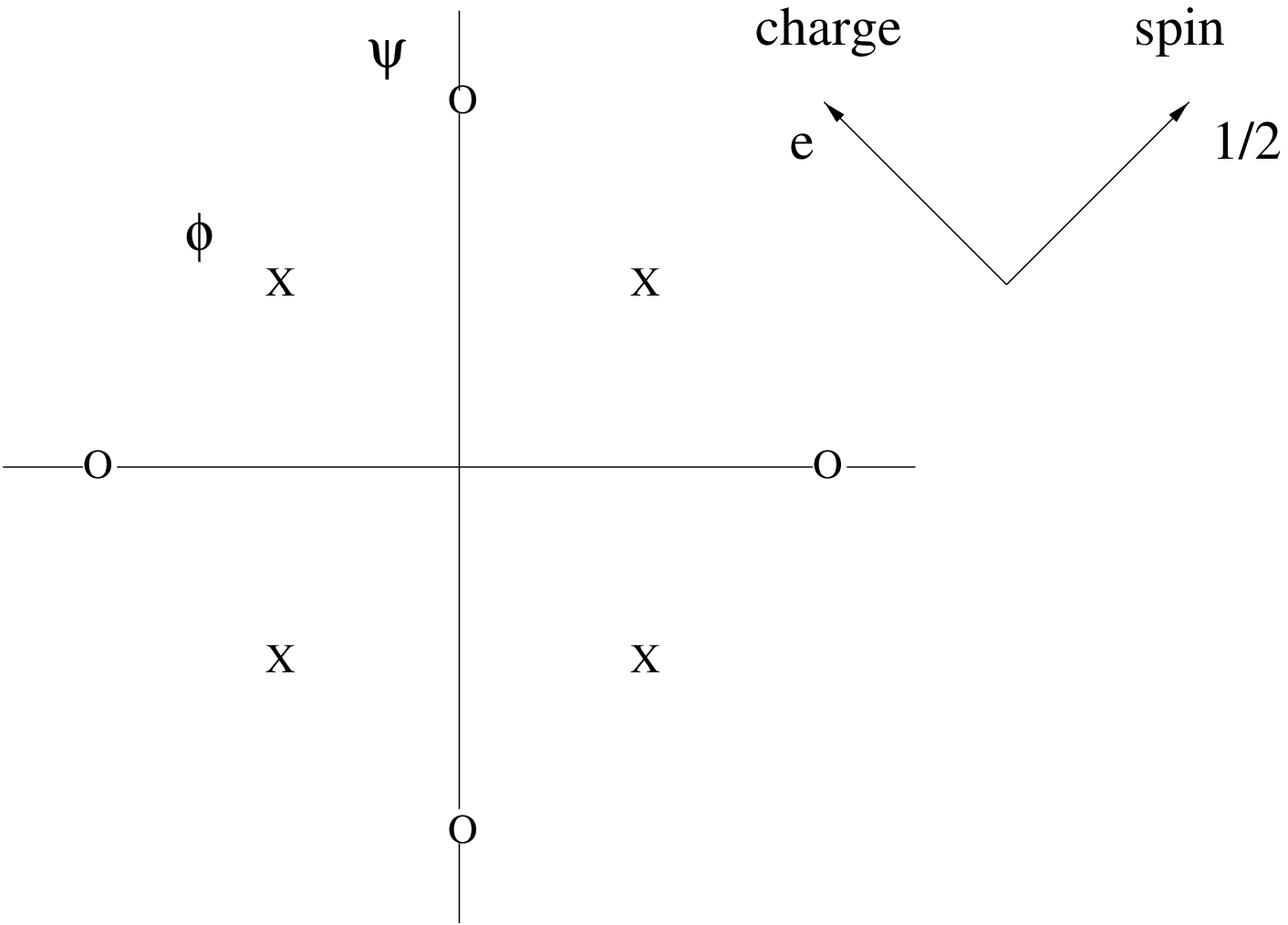}
\begin{figure}
\caption[]{$SO(4)$ weight diagram, showing the 
  quantum numbers of the vector (fermion) representation
  $\psi_{(\sigma,\pm e)}$ (open dots) and of the spinor 
  (spinon/holon) representation $(\phi_\sigma,\phi_{\pm e})$
  (crosses).}
\end{figure}
}

\vbox{
\epsfxsize=5.9cm
\epsfysize=4.5cm
\epsffile{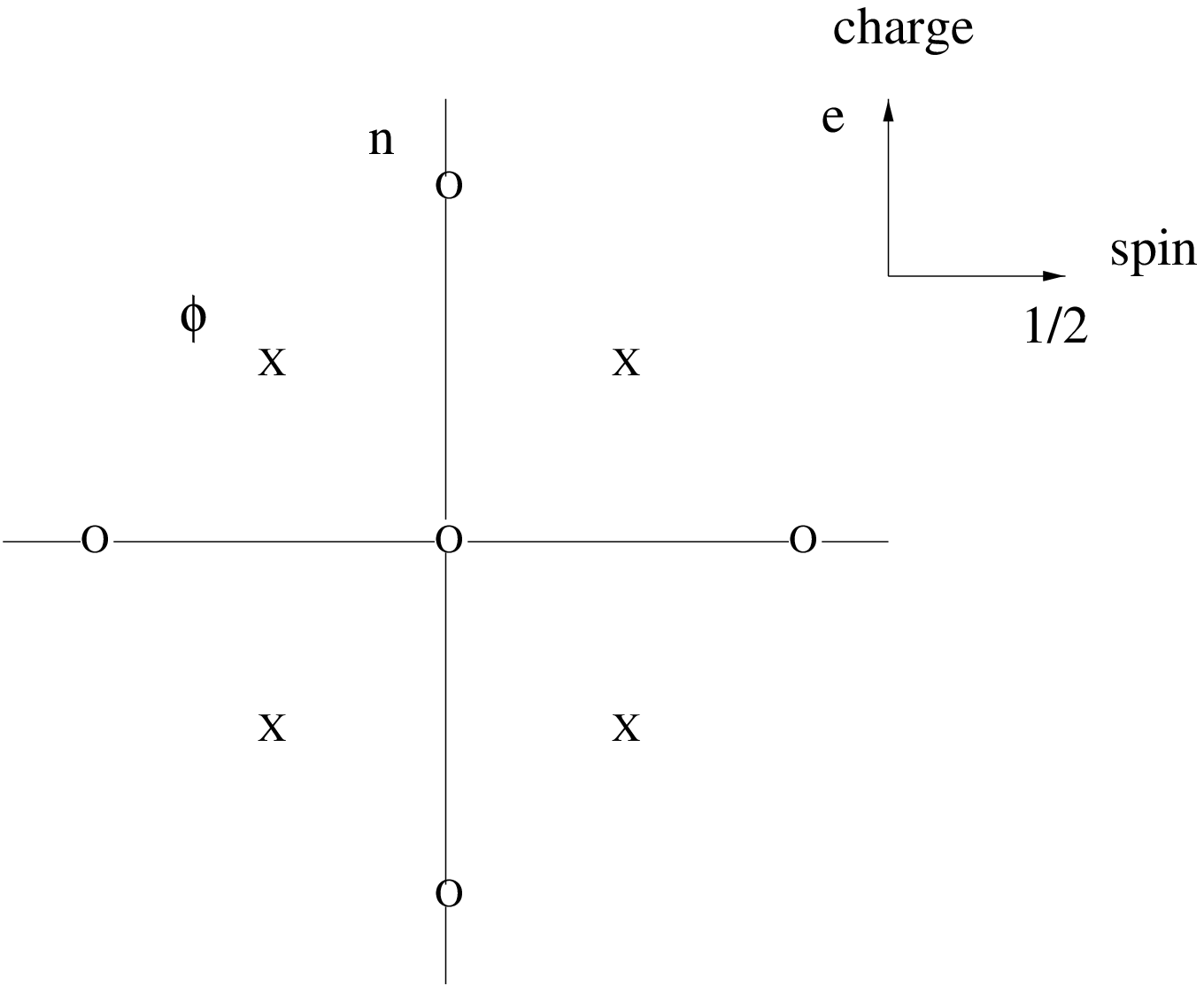}
\begin{figure}
\caption[]{$SO(5)$ weight diagram, showing the 
  quantum numbers of the vector (order parameter) representation
  $n_a$ (open dots) and of the spinor representation
  $\phi_{(\sigma,\pm e)}$ (crosses).}
\end{figure}
}

\end{document}